\documentclass[aps,twocolumn,prd,showpacs,amsmath,amssymb]{revtex4}
\usepackage{dcolumn}
\usepackage{bm}
\usepackage{amsfonts}
\usepackage{pifont}

\newcommand{\be}{\begin{equation}}
\newcommand{\ee}{\end{equation}} 
\newcommand{\eei}{\end{equation}\indent\indent}
\newcommand{\bc}{\begin{center}}
\newcommand{\ec}{\end{center}}
\newcommand{\ber}{\begin{eqnarray*}}
\newcommand{\ear}{\end{eqnarray*}}
\newcommand{\ba}{\begin{array}}
\newcommand{\ea}{\end{array}}

\newcommand{\hs}{\,-\,}

\newcommand{\bea}{\begin{eqnarray}}
\newcommand{\eea}{\end{eqnarray}}

\newcommand{\ei}{\end{itemize}}

\newcommand{\bra}[1]{\left(#1\right)}
\newcommand{\bras}[1]{\left[#1\right]}
\newcommand{\brac}[1]{\left\{#1\right\}}

\newcommand{\nab}{\nabla}

\newcommand \veps {\varepsilon} 

\newcommand{\lb}{\{}
\newcommand{\rb}{\}}
\newcommand{\A}{{\cal A}}
\newcommand{\E}{{\cal E}}
\renewcommand{\H}{{\cal H}}

\def\case#1/#2{\textstyle\frac{#1}{#2} }

\begin{document}

\title{Birkhoff Theorem and Matter}
\author{Rituparno Goswami and George F R Ellis}
\affiliation{ACGC and Department of Mathematics and Applied
Mathematics, University of Cape Town, Rondebosch, 7701, South
Africa}
\email{Rituparno.Goswami@uct.ac.za, George.Ellis@uct.ac.za}


\date{\today}
\begin{abstract}
Birkhoff's theorem for spherically symmetric vacuum spacetimes is a
key theorem in studying local systems in general relativity theory.
However realistic local systems are only approximately spherically
symmetric and only approximately vacuum.  In a previous paper, we
showed the theorem remains approximately true in an approximately
spherically symmetric vacuum space time. In this paper we prove the
converse case: the theorem remains approximately true in a
spherically symmetric, approximately vacuum space time.
\end{abstract}
\maketitle

\section{Birkhoff's Theorem}
Birkhoff's theorem (see e.g. \cite{HawEll73}) is a key theorem in
general relativity theory. It underlies the way local astronomical
systems decouple from the expansion of the universe. It states that
if a spacetime domain is locally (a) spherically symmetric and (b)
empty, then it necessarily has an extra symmetry: it is either
static or spatially homogeneous. That is, either the spacetime is
locally flat, or it is locally part of a Schwarzschild solution:
either the exterior part of a Schwarzschild solution outside the
event horizon (as in the solar system) or the part of the solution
inside the event horizon (as in collapse of a star to a
singularity).\\

The theorem actually applies to a somewhat wider class of solutions
than spherically symmetric spacetimes: it applies to all vacuum
locally rotationally symmetric (LRS) class II solutions, that is
vorticity-free solutions with a preferred spatial axis that are
invariant under rotations about that axis \cite{EllisLRS,EllisLRS1}.
We emphasize here that this is a local result: it does
not depend on boundary conditions at infinity.\\

However real astronomical systems are neither exactly spherically
symmetric, nor exactly empty. While it remains valid for the case of
an elecrovac solution (\cite{Din92}, section 18.1), Birkhoff's
theorem is not true in general when matter is present, as is shown
for example by the Lemaitre-Tolman-Bondi solutions \cite{LTB,Kra97}.
It remains true if the matter is static (\cite{CapFar11}, Section
4.3) but this will not be true in general. These results do not
include crucial cases such as the Solar System, which is neither
exactly empty nor exactly spherically symmetric.\\

In a previous paper \cite{GE} we showed that the result is stable to
small geometric perturbations: it remains true if spacetime is not
exactly spherically symmetric. Here we show that the result is
stable to small matter perturbations: it remains true if spacetime
is not exactly vacuum, as for example in the case of the solar
system.

\section{Birkhoff Theorem in LRS-II spacetimes}
We prove the result by using the 1+1+2 covariant formalism
\cite{Chris}. First we give a brief outline of the proof of the
exact result ~\cite{GE} and then the approximate result is a
straightforward generalization of the exact result using the 1+1+2
covariant perturbation theory.

\subsection{1+1+2 Covariant formalism}
In 1+3 covariant approach ~\cite{Covariant,Covariant1,Covariant2},
first we define a timelike congruence by a timelike unit vector
$u^a$ ($u^a u_a = -1$). Then the spacetime is {\it locally} split in
the form $R\otimes V$ where $R$ denotes the timeline along $u^a$ and
$V$ is the tangent 3-space perpendicular to $u^a$. Then  any vector
$X^a$ can be projected on the 3-space by the projection tensor
$h^a_b=g^a_b+u^au_b$. The vector $ u^{a} $ is used to define the
\textit{covariant time derivative} (denoted by a dot) for any tensor
$ T^{a..b}{}_{c..d} $ along the observers' worldlines defined by \be
\dot{T}^{a..b}{}_{c..d}{} = u^{e} \nab_{e} {T}^{a..b}{}_{c..d}~, \ee
and the tensor $ h_{ab} $ is used to define the fully orthogonally
\textit{projected covariant derivative} $D$ for any tensor $
T^{a..b}{}_{c..d} $ , \be D_{e}T^{a..b}{}_{c..d}{} = h^a{}_f
h^p{}_c...h^b{}_g h^q{}_d h^r{}_e \nab_{r} {T}^{f..g}{}_{p..q}~, \ee
with total projection on all the free indices.\\

In the (1+1+2) approach we further split the 3-space $V$, by
introducing {\bf a} 
spacelike unit vector $ e^{a} $ orthogonal
to $ u^{a} $ so that \be e_{a} u^{a} = 0\;,\; \quad e_{a} e^{a} = 1.
\ee Then the \textit{projection tensor} \be N_{a}{}^{b} \equiv
h_{a}{}^{b} - e_{a}e^{b} = g_{a}{}^{b} + u_{a}u^{b} -
e_{a}e^{b}~,~~N^{a}{}_{a} = 2~, \label{projT} \ee projects vectors
onto the tangent 2-surfaces orthogonal to $e^{a}$ \textit{and}
$u^a$, which, following  \cite{extension}, we will refer to as `{\it
sheets}'. Hence it is obvious that $e^aN_{ab} = 0 =u^{a}N_{ab}$. In
(1+3) approach any second rank symmetric 4-tensor can be split into
a scalar along $u^a$, a 3-vector, {a scalar part on the 3-space orthogonal to
$u^a$,}  and a {\it projected symmetric trace free} (PSTF) 3-tensor.
In (1+1+2) slicing, we can take this split further by splitting the
3-vector and PSTF 3-tensor with
respect to $e^a$. For example, in the 1+3 splitting,
the Energy Momentum Tensor $T_{ab}$ can be
written as
\be
T_{ab}=\mu u_au_b+q_au_b+u_aq_b+p h_{ab}+\pi_{ab}
\ee
Where the scalars $\mu=T_{ab}u^au^b$ and $p=(1/3)T_{ab}h^{ab}$ are the energy density
and isotropic pressure respectively. The 3-vector, $q^a=T_{cb}u^bh^{ca}$, is the heat flux
and the PSTF 3-tensor, $\pi_{ab}=T_{cd}h^c_{<a}h^d_{b>}$, defines the anisotropic stress.
In 1+1+2 splitting, we further split the fluid variables
$q^{a}$ and $\pi_{ab}$ as
\bea
q^{a} &=& Qe^{a} + Q^{a}~, \\
\pi_{ab} &=& \Pi\bras{ e_{a}e_{b} -\frac12 N_{ab}} + 2\Pi_{(a}e_{b)} + \Pi_{ab}~.
\label{Anisotropic}
\eea
The sheet carries a natural 2-volume element, the alternating Levi-Civita
2-tensor:
\be
\veps_{ab}\equiv\veps_{abc}e^{c} = \eta_{dabc}e^{c}u^{d}~, \label{perm}
\ee
where $\veps_{abc}$ is the 3-space permutation symbol the volume element of the
3-space and $\eta_{abcd}$ is  the space-time permutator or the 4-volume.\\

Now apart from the `{\it time}' (dot) derivative, of an object
(scalar, vector or tensor) which is the derivative along the
timelike congruence $u^a$, we now introduce two new derivatives,
which $ e^{a} $ defines, for any object $ \psi_{a...b}{}^{c...d}  $:
\bea \hat{\psi}_{a..b}{}^{c..d} &\equiv &
e^{f}D_{f}\psi_{a..b}{}^{c..d}~,
\\
\delta_f\psi_{a..b}{}^{c..d} &\equiv & N_{a}{}^{f}...N_{b}{}^gN_{h}{}^{c}..
N_{i}{}^{d}N_f{}^jD_j\psi_{f..g}{}^{i..j}\;.
\eea
The hat-derivative is the derivative along the $e^a$ vector-field
in the surfaces orthogonal to $ u^{a} $. The $\delta$ -derivative is the
projected derivative onto the sheet, with the projection on every free index.\\

We can now decompose the covariant derivative of $e^a$ in the direction orthogonal
to $u^a$ into it's irreducible parts giving
\be
{\rm D}_{a}e_{b} = e_{a}a_{b} + \frac{1}{2}\phi N_{ab} +
\xi\veps_{ab} + \zeta_{ab}~,
\ee
where
\bea
a_{a} &\equiv & e^{c}{\rm D}_{c}e_{a} = \hat{e}_{a}~, \\
\phi &\equiv & \delta_ae^a~, \\  \xi &\equiv & \frac{1}{2}
\veps^{ab}\delta_{a}e_{b}~, \\
\zeta_{ab} &\equiv & \delta_{\lb a}e_{b \rb }~.
\eea
We see that along the spatial direction $e_a$, $\phi$ represents the
\textit{expansion of the sheet},  $\zeta_{ab}$ is the \textit{shear of $e^{a}$}
(i.e. the distortion of the sheet) and $a^{a}$ its \textit{acceleration}.
We can also interpret $\xi$ as the \textit{vorticity} associated with $e^{a}$
so that it is a representation of the ``twisting'' or rotation of the sheet.
The other derivative of $e^a$ is its change along $u^a$,
\be
\dot{e_a}=\A u_a +\alpha_a,
\ee
where we have $\A=e^a\dot{u_a}$ and $\alpha_a=N_{ac}\dot{e^c}$.
Also we can write the (1+3) kinematical variables and Weyl tensor as follows
\bea
\Theta&=&h_a^b\nab_bu^a\\
\dot{u}^{a} &=& \A e^{a}+ \A^{a}~,\label{1+1+2Acc} \\
\omega^{a} &=& \Omega e^{a} +\Omega^{a}~,
\eea
\bea
\sigma_{ab} &=& \Sigma\bra{ e_ae_b - \frac{1}{2}N_{ab}} +
2\Sigma_{(a}e_{b)} + \Sigma_{ab}~, \\
E_{ab} &=& \E\bra{ e_{a}e_{b} - \frac{1}{2}N_{ab}} +
2\E_{(a}e_{b)} + \E_{ab}~, \\
H_{ab} &=& \H\bra{ e_{a}e_{b} - \frac{1}{2}N_{ab}} + 2\H_{(a}e_{b)} +
\H_{ab}~.\label{MagH}
\eea
where $E_{ab}$ and $H_{ab}$ are the electric and magnetic part of the
Weyl tensor respectively. Therefore the key variables of the 1+1+2 formalism are
\bea
\left[ \Theta, \A, \Omega,\Sigma, \E, \H, \phi,\xi,\mu, p, \Pi, Q,
\A^{a},\Omega^{a}, Q^a, \Pi^a,\right.\nonumber\\
\left. \Sigma^{a}, \alpha^a, a^a,
\E^{a}, \H^{a}, \Sigma_{ab}, \E_{ab}, \H_{ab},\zeta_{ab}, \Pi_{ab} \right] \,.
\eea
These variables (scalars , 2-vectors and PSTF 2-tensors)
form an {\it irreducible set} and completely describe a spacetime locally.\\

{Using} the above described (1+1+2) variables, the
full covariant derivatives of $e^{a}$ and $ u^{a}$ are \bea \nab_{a}
e_{b} &=& - \A u_{a}u_{b} - u_{a}\alpha_{b} + \bra{\Sigma +
\frac13\Theta} e_{a} u_{b} \nonumber\\&&+ \bra{ \Sigma_{a} -
\veps_{ac}\Omega^{c}} u_{b} + e_{a}a_{b} + \frac12\phi N_{ab}
\nonumber\\&&+ \xi\veps_{ab} + \zeta_{ab}~, \label{eFullDerive} \eea
\bea \nab_{a}u_{b} &=& -u_{a}\bra{\A e_{b} + \A_{b}} +
e_{a}e_{b}\bra{ \frac13\Theta + \Sigma } \nonumber\\&&+ e_{a}\bra{
\Sigma_{b} + \veps_{bc}\Omega^{c}} +
\bra{\Sigma_{a}-\veps_{ac}\Omega^c}e_{b} \nonumber\\&&+ N_{ab}\bra{
\frac13\Theta - \frac12\Sigma} + \Omega \veps_{ab} + \Sigma_{ab}~.
\label{uFullDerive} \eea

\subsection{Field equations for LRS-II spacetimes}

We know that for LRS-II spacetimes ~\cite{EllisLRS} (which are
rotation free Locally Rotationally Symmetric spacetimes) the 1+1+2
covariant variables $\brac{\A, \Theta,\phi, \Sigma,\E, \mu, p, \Pi,
Q }$ fully characterize the kinematics. The propagation, evolutions
and constraint equations for these variables {in such spaces}
are: \bea \hat\phi&= -&\frac12\phi^2+\bra{\frac13\Theta
+\Sigma}\bra{\frac23\Theta-\Sigma}\nonumber\\
&&-\frac23\mu -\frac12\Pi-\E\;, \label{hatphi}\\
\hat\Sigma-\frac23\hat\Theta&= -&\frac32\phi\Sigma-Q~,
\\
\hat\E-\frac13\hat\mu+\frac12\hat\Pi&= -&\frac32\phi\bra{\E+
\frac12\Pi}\nonumber\\
&&+ \bra{\frac12\Sigma-\frac13\Theta}Q~. \label{Ehat}
\eea
\bea
\dot\phi &= -&
\bra{\Sigma-\frac23\Theta}\bra{\A-\frac12\phi} +Q~,\label{Q}
\\
\dot\Sigma-\frac23\dot\Theta&= -&\A\phi +2\bra{\frac13\Theta
-\frac12\Sigma}^{2}\nonumber\\
&&+\frac13 \bra{\mu+3p}-\E +\frac12\Pi~,
\\
\dot\E-\frac13\dot\mu+\frac12\dot\Pi&= &\bra{\frac32\Sigma-\Theta}\E
+\frac14\bra{\Sigma-\frac23\Theta}\Pi\nonumber\\&&
+\frac12\phi Q-\frac12\bra{\mu+p}\bra{\Sigma-\frac23\Theta}~.
\label{Edot}
\eea
\bea
\dot\mu+\hat Q&= -&\Theta\bra{\mu+p}-\bra{\phi+2\A}Q -
  \frac32\Sigma\Pi~,
\\
\dot Q +\hat p+\hat\Pi&= -&\bra{\frac32\phi+\A}\Pi-
\bra{\frac43\Theta+\Sigma} Q \nonumber\\
&&-\bra{\mu+p}\A~,
\\
\hat\A-\dot\Theta&= -&\bra{\A+\phi}\A+\frac13\Theta^2 \nonumber\\
&&   +\frac32\Sigma^2 + \frac12\bra{\mu+3p}~. \label{ray1} \eea

Since the vorticity vanishes, the unit vector field $ u^{a} $ is
hypersurface-orthogonal to the spacelike 3-surfaces whose intrinsic
curvature can be calculated from the \textit{Gauss equation} for $
u^{a} $ that is generally given as \cite{Gary}: \be
^{(3)}R_{abcd}=\bra{R_{abcd}}_{\bot} - K_{ac}K_{bd} +
K_{bc}K_{ad}\;, \ee where $ ^{(3)}R_{abcd} $ is the
\textit{3-curvature tensor}, $ \bot $ means projection with $ h_{ab}
$ on all indices and $ K_{ab} $ is the \textit{ extrinsic
curvature}. With the additional constraint of the vanishing of the
sheet distortion $ \xi $, {\it i.e.} the sheet is a genuine
2-surface. The Gauss equation for $ e^{a} $ together with the
3-Ricci identities determine the 3-Ricci curvature tensor of the
spacelike 3-surfaces orthogonal to $ u^{a} $ to be \be ^{3}R_{ab} =
-\bras{\hat{\phi}+\frac12 \phi^{2}}e_{a}e_{b} - \bras{\frac12
\hat{\phi} + \frac12\phi^{2} - K}N_{ab}\;, \ee This gives the
3-Ricci-scalar as \be ^{3}R = -2\bras{\hat{\phi} + \frac34\phi^{2} -
K} \label{3nRicci} \ee where $ K $ is the \textit{Gaussian
curvature} of the sheet, $ ^{2}R_{ab}=KN_{ab} $ . From this equation
and (\ref{hatphi}) an expression for $ K $ is obtained in the form
\cite{Gary} \be K = \frac13 \mu - \E - \frac12 \Pi + \frac14
\phi^{2} - \bra{\frac13 \Theta - \frac12 \Sigma}^{2}
\label{GaussCurv} \ee From (\ref{hatphi}-\ref{Edot}), the evolution
and propagation equations of $K$ can be determined as \bea
\dot{K} = - \bra{\frac23 \Theta - \Sigma}K,\label{evoGauss}\\
\hat{K} = -\phi K. \label{propGauss} \eea From equation
(\ref{evoGauss}), it follows that whenever the Gaussian curvature of
the sheet is non-zero and constant in time, then the shear is always
proportional to the expansion as $ \Sigma=\frac23 \Theta $.

\subsection{Vacuum LRS-II spacetimes and Birkhoff Theorem}

To covariantly investigate the geometry of the vacuum LRS-II
spacetime, we write the {\it Killing equation} for a
vector of the form
\be \xi_a=\Psi u_a+\Phi e_a, \label{KV}\ee
where $\Psi$
and $\Phi$ are scalars. The Killing equation gives \be \nab_a(\Psi
u_b+\Phi e_b) + \nab_b(\Psi u_a+\Phi e_a) =0\;. \label{Killing} \ee
which in this case becomes the following differential equations and
constraints:
\bea
\dot\Psi+\A\Phi&=&0, \label{psidot}\\
\hat\Psi -\dot\Phi-\Psi\A+\Phi(\Sigma+\frac13\Theta)&=& 0,\label{psihat}\\
\hat\Phi+\Psi(\frac13\Theta+\Sigma)&=&0,\label{cons1}\\
\Psi(\frac23\Theta-\Sigma)+\Phi\phi&=&0.\label{cons2} \eea Now we
know $\xi_a\xi^a=-\Psi^2+\Phi^2$. If $\xi^a$ is timelike (that is
$\xi_a\xi^a<0$), then because of the arbitrariness in choosing the
vector $u^a$, we can always make $\Phi=0$. On the other hand, if
$\xi^a$ is spacelike (that is  $\xi_a\xi^a>0$), we can make
$\Psi=0$.\\

Let us assume that $\xi^a$ is timelike and set $\Phi=0$. In that
case Killings equations (\ref{psidot}-\ref{cons2})become
\bea
\dot\Psi&=&0, \label{psidot1}\\
\hat\Psi -\Psi\A &=& 0,\label{psihat1}\\
\Psi(\frac13\Theta+\Sigma)&=&0,\label{cons11}\\
\Psi(\frac23\Theta-\Sigma)&=&0.\label{cons21} \eea We know that the
solution of equations (\ref{psidot1}) and (\ref{psihat1}) always
exists while the constraints (\ref{cons11}) and (\ref{cons21})
together imply that in general, (for a non trivial $\Psi$),
$\Theta=\Sigma=0$. When these are plugged into the field equations
(\ref{hatphi})-(\ref{ray1}), we see that the ``dot'' derivative of
all the quantities vanish and the remaining field equations are as
follows: \bea
\hat\phi&= -&\frac12\phi^2-\E, \label{FE1}\\
\hat\E&=-&\frac32\phi\E\,
\label{FE2}\\
\E&=&-\A\phi, \label{FE3}\\
\hat\A&=-&(\A+\phi)\A\;. \label{FE4}\\
\eea
Also the local Gaussian curvature of the 2-sheets are given as
\be
K = - \E + \frac14 \phi^{2}
\label{GaussCurv1}
\ee
From ~\cite{GE} we know that the resultant set of equations has
a unique solution (for $K>0$), which gives the Schwarzschild metric.
Similarly if the Killing vector is spacelike we have $\A=\phi=0$. In
that case the spacetime is spatially homogeneous as the `hat'
derivative of all the quantities vanish and the resultant solution
(for $K>0$) is the Schwarzschild interior.\\

Hence the Birkhoff Theorem for LRS-II spacetime says that there
always exists a Killing vector in the local $[u,e]$ plane
for a vacuum LRS-II spacetime. If the Killing vector is timelike then the spacetime is locally
static, and if the Killing vector is spacelike the spacetime is locally spatially
homogeneous. For $K>0$, we get the known result, any
$C^2$ solution of Einstein's equations in empty space which is
spherically symmetric in an open set ${\mathcal{S}}$ is locally
equivalent to part of maximally extended Schwarzschild solution in
${\mathcal{S}}$.\\

Also from ~\cite{GE}, we know that for vacuum LRS-II spacetime \be
\E=CK^{3/2}. \label{EK} \ee That is, the 1+1+2 scalar of the
electric part of the Weyl tensor is always proportional to a power
of the Gaussian curvature of the 2-sheet. The proportionality
constant $C$ sets up a scale in the problem. We can immediately see
that for Minkowski spacetime $C=0$. Also it is interesting to note
that the modulus of the proportionality constant in equation
(\ref{EK}), is exactly equal to the Schwarzschild radius: \be C =
R_S = 2M \label{EK1} \ee where $M$ is the mass of the star in the
unit of $8\pi G=c=1$.

\section{Almost Vacuum LRS-II spacetimes}

The result obtained in the previous section is not true if spacetime
is not a vacuum (empty) spacetime, for the degrees of freedom
available through a matter source generically invalidate the result,
as is shown for example by the family of Lema\^{\i}tre-Tolman-Bondi
(LTB) models \cite{LTB}. However we would like to ask the question,
that how much matter can be present if the above theorem is to
remain approximately true. In other words, we would like to perturb
a vacuum LRS-II spacetime by introducing a small amount of general
matter in the spacetime. In this section we only deal with the
static exterior background as that is astrophysically more
interesting.

\subsection{Matter} We know from the covariant linear perturbation
theory, any quantity which is zero in the background is considered
as the first order quantity and is automatically gauge-invariant by
virtue of the Stewart and Walker lemma \cite{SW}. Hence the set
$\brac{\Theta,\Sigma,\mu, p, \Pi, Q }$, describes the first order
quantities. As we have already seen that the vacuum spacetime has an
covariant scale given by the the Schwarzschild radius which sets up
the scale for perturbation. Let us locally introduce general matter
on a static Schwarzschild background such that \be \left[
\frac{\mu}{K^{(3/2)}}, \frac{|p|}{K^{(3/2)}},
\frac{|\Pi|}{K^{(3/2)}}, \frac{|Q|}{K^{(3/2)}}\right]<< C,
\label{cond1} \ee and \be \left[ \frac{|\hat\mu|}{K^{(3/2)}},
\frac{|\hat p|}{K^{(3/2)}}, \frac{|\hat\Pi|}{K^{(3/2)}} \frac{|\hat
Q|}{K^{(3/2)}}, \right]<< \phi C \label{cond2} \ee where $C$ is the
proportionality constant of (\ref{EK}), which is also the
Schwarzschild radius.

\subsection{Domains} Now we need to make clear in what domain these
equations will hold. The application will be to the spherically
symmetric exterior domain of a star of mass $M$ and Schwarzschild
radius $R_S = 2M$, in the units of $8\pi G=c=1$. We will define
\emph{Finite Infinity} ${\cal F}$ as a 2-sphere of radius $R_{\cal
F} \gg R_M$ surrounding the star: this is infinity for all practical
purposes \cite{Ell84,StoEll}. We assume the relations (\ref{cond1},
\ref{cond2}) hold in the domain $D_{\cal F}$ defined by $r_S < r <
R_{\cal F}$ where $r_S > r_M $ is the radius of the surface of the
star. This is the local domain where our results will apply. In the
case of the solar system, $R_{\cal F}$
can be taken to be about a light year {(we return to this issue in Section 4)}.\\

It is important to make this restriction, else eventually we will
reach a radius $r$ where these inequalities may no longer hold; but
this will be unphysical, as in the real universe asymptotically flat
regions are always of finite size, being replaced at larger scales
by galactic and cosmological conditions. The result we wish to prove
is a local result, applicable to the locally restricted nature of
real physical systems.

\subsection{Equations} Now subtracting the background equations
(\ref{FE1})-(\ref{FE4}), from the field equations
(\ref{hatphi})-(\ref{ray1}), and neglecting the higher order
quantities, we get the following linearised equations for the first
order quantities \be
\hat\Sigma-\frac23\hat\Theta=-\frac32\phi\Sigma-Q~,\label{LE1} \ee
\be \dot\Theta= -\frac12\bra{\mu+3p}~.\label{LE8} \ee \be
\dot\Sigma-\frac23\dot\Theta=\frac13
\bra{\mu+3p}+\frac12\Pi~,\label{LE4} \ee \be
\dot\phi=\bra{\Sigma-\frac23\Theta}\bra{\A-\frac12\phi}
+Q~,\label{LE3} \ee \be
\frac13\hat\mu-\frac12\hat\Pi=\frac34\phi\Pi~, \label{LE2} \ee \be
\dot\E-\frac13\dot\mu+\frac12\dot\Pi=\bra{\frac32\Sigma-\Theta}\E
+\frac12\phi Q~,\label{LE5} \ee \be \dot\mu+\hat
Q=-\bra{\phi+2\A}Q~, \label{LE6} \ee \be \dot Q +\hat p+\hat\Pi=
-\bra{\frac32\phi+\A}\Pi-\bra{\mu+p}\A~,\label{LE7} \ee

Equations (\ref{LE2})-(\ref{LE7}) are linearised matter conservation
equations. From these equations we can see that if (\ref{cond1}) and
(\ref{cond2}) are locally satisfied at any epoch, within the domain
$D_{\cal F}$, then the time
variation of the matter variables are of same order of smallness as
themselves. Hence there exists an open set $\mathcal{S}$ within where the
amount of matter remains ``{\it small}'', if the amount is small at any
epoch in $\mathcal{S}$ and only small amounts of matter
enter $D_{\cal F}$ across ${\cal F}$. One could attempt to determine
the same kinds of inequality as those above for matter crossing
${\cal F}$, but one can resolve this issue in another way: we have
not yet specified the time evolution of ${\cal F}$. We now do so in
the following manner: choose it in a suitable manner in some initial
surface $t = t_0$, and then propagate it to the future by dragging
it along world lines that are integral curves of the timelike
eigenvector of the Ricci tensor $R_{ab}$ (this will be unique for
any realistic non-zero matter). As these are then timelike
eigenvectors of the stress tensor $T_{ab}$ (because of the field
equations), equal amount of energy density will convect in and out
across ${\cal F}$ due to random motions of matter \cite{Covariant1};
the total amount of matter inside ${\cal F}$ will be conserved, and
if the inequalities (\ref{cond1}, \ref{cond2}) are satisfied at some
initial time they will be satisfied at later times, unless major
masses enter the ${\cal F}$ locally in some region. If this is so,
we do not have an isolated system and the extended Birkhoff's theorem need not apply.\\

Hence we will define the time evolution of ${\cal F}$ in the way
just indicated, and suppose that (\ref{cond1}, \ref{cond2}) are then
satisfied at later times; if this is not the case the local system
considered is not isolated and our result is not applicable.

\subsection{Almost symmetries} Now from equations
(\ref{LE1})-(\ref{LE4}), it is evident that if the matter variables
remains ``{\it small}'' as defined by (\ref{cond1}) then the spatial
and temporal variance of the expansion $\Theta$ and the shear
$\Sigma$ are of the same order of smallness as the matter. In that
case we see that a timelike vector will not exactly solve the
Killing equations (\ref{psidot})-(\ref{cons2}) in general, although
it may do so approximately. To see this explicitly, let us set $\Phi
= 0$ in (\ref{KV}) and consider the following symmetric tensor \be
K_{ab} := \nab_a(\Psi u_b) + \nab_b(\Psi u_a) \;. \label{Killing1}
\ee This tensor vanishes if $\Psi u^a$ is a Killing vector. This is
the case of an exact symmetry when the spacetime is exactly static.
However, in the perturbed scenario, to see how close the vector
$\xi_a = \Psi u_a$ is to a Killing vector, let us consider the
scalars constructed by contracting the above tensor by the vectors
$u^a$, $e^a$ and the projection tensor $N^{ab}$. If the conditions
\be
\left[\frac{|K_{ab}u^au^b|^2}{K^{3/2}},\frac{|K_{ab}u^ae^b|^2}{K^{3/2}},
\frac{|K_{ab}e^ae^b|^2}{K^{3/2}},
\frac{|K_{ab}N^{ab}|^2}{K^{3/2}}\right]<<C \label{cond3} \ee are
satisfied, then we can say that $\xi_a = \Psi u_a$ is close to a
Killing vector and the spacetime is approximately static.

\subsection{The Main Result} From equations (\ref{eFullDerive}) and
(\ref{uFullDerive}), we see that there always exists a non-trivial
solution of the scalar $\Psi$ for which $|K_{ab}u^au^b|$ and
$|K_{ab}u^ae^b|$ vanishes; we choose $\Psi$ accordingly. However for
a general matter perturbation, as $\Theta$ and $\Sigma$ are
non-zero, from (\ref{eFullDerive}) and (\ref{uFullDerive}) it is
evident that $|K_{ab}e^ae^b|^2$ and $|K_{ab}N^{ab}|^2$ are non-zero.
However, subtracting the background equation (\ref{GaussCurv1}) from
(\ref{GaussCurv}), we get \be \bra{\frac13 \Theta - \frac12
\Sigma}^{2}~\approx  \frac13 \mu- \frac12 \Pi. \label{KE1} \ee
Similarly subtracting (\ref{FE1})from (\ref{hatphi}) we get \be
\bra{\frac13\Theta+\Sigma}\bra{\frac23\Theta-\Sigma}~\approx
\frac23\mu +\frac12\Pi\;. \label{KE2} \ee Using the above equations
(\ref{KE1}) and (\ref{KE2}), we immediately see that if the amount
of matter is ``small'', that is the condition (\ref{cond1}) is
satisfied, then the following conditions are satisfied \be
|K_{ab}e^ae^b|^2=\Psi^2(\frac13\Theta+\Sigma)^2\ll C K^{3/2},
\label{cons12} \ee \be
|K_{ab}N^{ab}|^2=\Psi^2(\frac23\Theta-\Sigma)^2\ll C K^{3/2}.
\label{cons22} \ee Therefore we can say that there always exists a
timelike vector that satisfies (\ref{cond3}). This vector then
almost solves the Killing equations in $\mathcal{S}$ and hence the
spacetime is {\it almost} static in $\mathcal{S}$. Also the
resultant field equations are the zeroth order equations
(\ref{FE1})-(\ref{FE4}) with $\mathcal{O}(\epsilon)$ terms added to
it. Hence for $K>0$, the local spacetime is described by an {\it
almost} Schwarzschild
metric.\\

The above conditions, (\ref{cond1}) and (\ref{cond2}), can also be
written in another way. \be \left[ \frac{|R|}{K^{(3/2)}},
\frac{|R_{ab}u^au^b|}{K^{(3/2)}},
\frac{|R_{<ab>}e^ae^b|}{K^{(3/2)}},
\frac{|R_{<ab>}u^ae^b|}{K^{(3/2)}} \right]<< C \label{cond4} \ee and
\be \left[ \frac{|\hat
R|}{K^{(3/2)}},\frac{|R_{ab}u^au^b|\hat{}}{K^{(3/2)}},
\frac{|R_{<ab>}u^ae^b|\hat{}}{K^{(3/2)}},\frac{|R_{<ab>}e^ae^b|\hat{}}
{K^{(3/2)}}\right]<< \phi C
\label{cond5} \ee 
In other words the ratio of the scalars
constructed from the Ricci tensor using the vectors $u^a$ and $e^a$
(and their spatial variations) to the $(3/2)$th power of the local
Gaussian curvature of the 2-sheet should be much smaller than the
Schwarzschild radius if the Birkhoff theorem is to remain
approximately true. Equations (\ref{cond4}) and (\ref{cond5}) are
easier to use, in case of presence of multifluids in the spacetime.

\section{Comments on the solar system}

In case of the solar system  ~\cite{SS} we know that the
interplanetary medium includes interplanetary dust, cosmic rays and
hot plasma from the solar wind. Its density is very low at about 5
particles per cubic centimeter in the vicinity of the Earth; it
decreases with increasing distance from the sun, in inverse
proportion to the square of the distance. In this section, to
compare our result with the observed astronomical data, we will use
SI units for clarity.\\

The density of interplanetary medium is variable, and may be
affected by magnetic fields and events such as coronal mass
ejections. It may rise to as high as 100 particles/$cm^3$. These
particles are mostly Hydrogen nuclei, and hence the maximum density
per cubic meter will be approximately of the order of $10^{-19}$
Kilograms, and the local Gaussian curvature of the heliocentric
celestial sphere in the vicinity of the earth is of the order of
$10^{-22}\;\;m^{-2}$ . Hence the ratio of the maximum interplanetary
density to the  $(3/2)$th power of the local Gaussian curvature is
of the order of $10^{14}$ Kilograms, which is much smaller then the
solar mass ($10^{30}$ Kilograms). Also the large amplitude waves in
the medium are comparable to the energy density of the unperturbed
medium, which makes the spatial variation of energy density to be of
the same order of smallness as itself. This satisfies (\ref{cond1})
and (\ref{cond2}) and hence in the solar
system the Birkhoff theorem remains almost true.\\

We can relate { the discussion} to the \emph{Finite Infinity}
concept { for the solar system}. We know that the outer edge of
the solar system is the boundary between the flow of the solar wind
and the diffused interstellar medium. This boundary, which is known
as the {\it Heliopause}, is at a radius of approximately $10^{13}$
meters. The interplanetary medium thus fills the roughly spherical
volume contained within the heliopause. As the density of the
interplanetary medium decreases in inverse proportion to the square
of the distance, the density near the heliopause is of the order of
$10^{-23}$ Kilograms per cubic meter.  Hence the ratio of the
density to the  $(3/2)$th power of the local Gaussian curvature is
of the order of $10^{16}$ Kilograms and still remains much smaller
than the solar mass. Also the amount of matter crossing the
heliopause to the diffused interstellar medium is of the same order.
Hence we can easily define the heliopause as the boundary of our
domain $D_{\mathcal{F}}$. As the conditions
 (\ref{cond1}) and (\ref{cond2}) { are} 
 true at the boundary of the domain,
they should be true everywhere inside the domain, unless the matter
outside the star is highly clustered locally. But we are considering
the case 
{of} a low density diffuse gas where this is not the
case. {the conditions (\ref{cond4}) and (\ref{cond5}) will be satisfied in this domain.}\\

For the massive planets inside the solar system (e.g. Jupiter or
Saturn), these conditions may be violated in their very close
vicinity, but in that case the local spacetime no longer remains
spherically symmetric. However as the vast fraction of the solar
system's mass (more than $99\%$) is in the sun, on average these
massive planets have a very tiny effect on the system as a whole and
the approximate theorem remains true. Hence the local spacetime
within the solar system is ``{\it almost}'' described by a
Schwarschild metric.

\section{Conclusion} Our previous paper showed an ``Almost Birkhoff
theorem'' holds if a vacuum spacetime is almost spherically
symmetric. This paper shows such a result also holds for an almost
vacuum spherically symmetric spacetime. \\

It seems clear that the generic result {-- needed for the real
universe application --} will be true: an ``Almost Birkhoff
theorem'' will hold for an almost--vacuum almost--spherically
symmetric spacetime. We leave that proof, {combining the results
of this paper and the previous one, for another investigation}.

\end{document}